\documentclass[
	showpacs,
	aps,
	pra,
	preprint,
	superscriptaddress,
	longbibliography
]{revtex4-1}
\usepackage{amsmath}
\usepackage[draft=false]{hyperref}
\usepackage{mathtools}
\usepackage[utf8]{inputenc}
\usepackage{graphicx}
\usepackage{braket}

\newcommand{\bk}{\mathbf{k}}
\newcommand{\bp}{\mathbf{p}}
\newcommand{\bq}{\mathbf{q}}
\newcommand{\dd}{\mathrm{d}}

\begin{document}

\title{\textsc{Plasmonic Cooper pairing in single layer graphene}}
\date{\today}
\author{D. Elst}
\email{Dietrich.Elst@uantwerpen.be}
\affiliation{TQC, Universiteit Antwerpen, Universiteitsplein 1, B-2610 Antwerpen, Belgium}
\author{S. N. Klimin}
\affiliation{TQC, Universiteit Antwerpen, Universiteitsplein 1, B-2610 Antwerpen, Belgium}
\author{J. Tempere}
\affiliation{TQC, Universiteit Antwerpen, Universiteitsplein 1, B-2610 Antwerpen, Belgium}
\affiliation{Lyman Laboratory of Physics, Harvard University, Cambridge, MA 02138, USA}

\begin{abstract}
    The dielectric function method (DFM), which uses a non-adiabatic approach to calculate the critical temperatures for superconductivity, has been quite successful in describing superconductors at low carrier densities.
    This regime of carrier densities causes other theories, such as BCS and Migdal-Eliashberg theory, to violate their assumption of a small Debye window.
    We investigate the application of DFM to the linear dispersion of single layer graphene.
    We derive the gap equation of DFM for a Dirac cone and calculate the critical temperature as a function of carrier density.
    This is done using an interaction potential that utilizes the Random Phase Approximation dielectric function and thus allows for plasmonic interactions.
    Our results show a significantly different behaviour of the critical temperature as a function of carrier density when compared to the BCS result.
    Thus, we find the DFM approach to be better suited when considering graphene systems at low carrier densities.
\end{abstract}
\maketitle

\section{\label{sec:intro}Introduction}
The two-dimensional honeycomb structure of single layer graphene has raised substantial interest since its experimental realization \cite{novoselov04}.
Several unusual electronic properties are supported by the symmetry of graphene's two-dimensional electron gas (2DEG). 
Examples of these effects are the half-integer quantum Hall effect observable at high temperature \cite{novoselov05, zhang05}, and conductivity at zero doping \cite{novoselov05}. 
Combined with its strength, flexibility \cite{lee08, briggs10}, and potential as a building block in creating composite materials, graphene is a promising material for the development of technological advancements. 
By applying an electric field, the chemical potential of graphene can be tuned to lie above or below the Dirac point.
This allows for a setup that is able to control the type and density of charge carriers by the application of a voltage.

Graphene samples have been shown to support the propagation of Cooper pairs through the proximity effect \cite{heersche07, du08, shalom16}.
More recently, intrinsic superconductivity has been observed in twisted bilayer graphene \cite{cao18}.
Theoretically, anomalous Andreev reflection has been predicted for graphene-superconductor junctions \cite{beenakker06}.
For undoped graphene, Kopnin and Sonin~\cite{kopnin08} predicted the presence of a critical point in the interaction strength below which the critical temperature vanishes.
However, they also demonstrated that Cooper pairing is possible for finite doping at arbitrary coupling strengths.
These predictions were made using a general BCS model that does not select a specific pairing mechanism.
Uchoa and Neto~\cite{uchoa07} investigated superconductivity in metal coated graphene with a BCS model.
They found the electron-plasmon mechanism of superconductivity to be favorable at low electron doping densities.
However, their proposed technique to achieve the relevant doping densities is through adatoms, which introduces additional screening effects that are not present in a single isolated layer of graphene.
For the interesting system of twisted bilayer graphene, Ref.~\cite{sharma19} shows the plasmon mechanism is one of the key processes facilitating the superconducting phase transition of the system.

At low doping theoretical descriptions of superconductivity that are perturbative in the size of the interaction region with regard to the Fermi level, such as BCS or Migdal-Eliashberg theory, lose accuracy.
An approach that is able to include a broader interaction region, such as the Dielectric Function Method (DFM), is more appropriate in this case.
As far as we know, the DFM technique was never applied to study superconductivity in weakly doped graphene. As we will show, the results  deviate from those of the standard BCS approach.

First introduced by Kirzhnits et al. \cite{kirzhnits73}, later refined by Takada \cite{takada78} and recently verified by Rosenstein et al.~\cite{rosenstein16}, DFM uses the dielectric function to describe screening effects in the weak-coupling regime.
This way, a general form of the electron-electron interaction can be used that is not limited to a small interaction window around the Fermi level.
This technique has already been successfully applied to systems such as bulk SrTiO$_3$ \cite{klimin17,klimin19} and the 2DEG at the LaAlO$_3$-SrTiO$_3$ interface \cite{klimin14,klimin17, rosenstein16}.

In this paper, we apply the DFM technique to single layer graphene at low electron doping and compare the results with those of standard BCS theory.
In Section \ref{sec:theory}, we review the DFM technique and construct a relevant dielectric function in the Random Phase Approximation (RPA) suitable to describe plasmonic interactions.
Using the DFM method, we investigate plasmon mediated pairing \cite{uchoa07} and the modification of the critical temperature by the dielectric constant in Section \ref{sec:results}.
We compare our results with the BCS model of Ref.~\cite{kopnin08}.

\section{\label{sec:theory}Theory}
\subsection{\label{subsec:DFM}DFM equations}
Graphene's electronic band structure is well described by the tight binding Hamiltonian
\begin{equation*}
	\hat{H} = - t \sum_{\langle i,j \rangle} \left( \hat{c}_i^\dagger \hat{c}_j + \hat{c}_j^\dagger \hat{c}_i \right),
\end{equation*}
where $t \approx 2.8$ eV is the hopping matrix element, $\hat{c}_i$ the annihilation operator for an electron at site $i$ of the hexagonal lattice, and the summation only includes nearest neighbour hopping.
Two points of interest are the $K$ and $K'$ symmetry points.
Here, the valence and conduction bands touch with a linear dispersion.
Thus, for small Fermi levels, the electronic energy dispersion is given by the equation
\begin{equation}
    \label{eq:dirac}
    \epsilon_{\lambda,\bk} = \lambda v_F |\bk| - \epsilon_F,
\end{equation}
which contains an offset to position its zero at the Fermi level $\epsilon_F$, rather than at the Dirac point.
In all equations, we use units where $\hbar = k_B = 1$.
The parameter $\lambda$ denotes the conduction ($\lambda = 1$) or valence band ($\lambda = -1$), while the constant $v_F = 3ta/2$ is the Fermi velocity with $a$ the bond length between two neighbouring carbon atoms.

The linear dispersion of graphene and its two-dimensional nature will lead to a different gap equation than the original DFM gap equations derived by Kirzhnits et al.~\cite{kirzhnits73} and Takada~\cite{takada78} for materials with a parabolic band dispersion.
We start from the general weak-coupling expression for the superconducting gap $\Delta(\bp)$ at a temperature $T \lesssim T_c$, given by Takada~\cite{takada78}:
\begin{widetext}
\begin{equation}
	\label{eq:takgap}
	\Delta(\bp) = - \sum_{\bk} \frac{\Delta(\bk)}{2|\epsilon_{\lambda,\bk}|}
	\tanh\left( \frac{|\epsilon_{\lambda,\bk}|}{2T_c} \right)
	\frac{2}{\pi} 
	\int \limits_0^\infty \dd\Omega \frac{
	\vert \epsilon_{\lambda,\bk} \vert+\vert \epsilon_{\lambda,\bp} \vert }{\Omega^2 + \left( \vert \epsilon_{\lambda,\bk} \vert+ \vert \epsilon_{\lambda,\bp} \vert  \right)^2}V(\bk + \bp, i\Omega),
\end{equation}
\end{widetext}
where $V(\bk + \bp, i\Omega)$ is the effective electron-electron interaction potential.
For a full anisotropic treatment, this equation can prove difficult to solve, even numerically.
For a good convergence of the solution $\Delta(\bp)$, the summation over $\bk$ must be sampled relatively fine over a large region, which induces a large computational load.
Equation (\ref{eq:takgap}) can be simplified for a sample of graphene at low electron doping.
Indeed, at carrier concentrations below $10^{12}$~cm$^{-2}$, the Fermi level is sufficiently small ($\lesssim 0.1$ eV) for the Dirac cone to be a good approximation of the graphene band structure.
This linear isotropic dispersion reduces the complexity of the gap equation.

After using the dispersion given in Eq.~(\ref{eq:dirac}) and converting the summation over $\bk$ to an integration, the gap equation becomes
\begin{equation}
	\Delta(\omega) = - \int_{-\epsilon_F}^\infty \dd\omega' \frac{\Delta(\omega')}{2\omega'} \tanh \left( \frac{\omega'}{2T_c} \right)  K(\omega, \omega'),
	\label{eq:gapequation}
\end{equation}
where the kernel $K(\omega,\omega')$ is defined as
\begin{equation}
	K(\omega, \omega') = \frac{\omega' + \epsilon_F}{\pi^3v_F^2} \int_0^\pi \dd\theta \int_0^\infty \dd\Omega \frac{|\omega| + |\omega'|}{\Omega^2 + \left( |\omega|+ |\omega'|  \right)^2}V(\bq, i\Omega),
	\label{eq:kernel}
\end{equation}
with the vector $|\bq| = |\bp + \bk| = \sqrt{p^2 + k^2 + 2k p \cos\theta}$ and the (isotropic) energies $\omega = \epsilon_{+\bp}$ and $\omega' = \epsilon_{+\bk}$. 
The two-dimensional result by Takada~\cite{takada78} differs solely in the prefactor of the kernel.
Equation (\ref{eq:gapequation}) is very similar in structure to the BCS gap equation \cite{bardeen57}
\begin{equation*}
	\Delta = V \sum_{\bk \in \mathcal{D}} \frac{|\Delta|}{2\epsilon_\bk} \tanh\left( \frac{\epsilon_\bk}{2 T_c} \right),
\end{equation*}
where $V$ is the BCS interaction strength, $\mathcal{D}$ is the Debye window, and $\epsilon_\bk$ is the dispersion.
Equation~(\ref{eq:gapequation}) does not contain the interaction potential directly, it is incorporated in the kernel $K(\omega,\omega')$. 
However, when the interaction potential $V(\bq,i\Omega)$ of Eq.~(\ref{eq:kernel}) is chosen to be
\begin{equation*}
	V(\bq,i\Omega) = \left\{
	\begin{aligned}
			-V \qquad &\bk \in \mathcal{D}\\
			0 \qquad &\text{elsewhere}
	\end{aligned}
	\right.,
\end{equation*}
the DFM gap equation simplifies to exactly the BCS gap equation.
Migdal's theorem~\cite{migdal58} shows that, by introducing the window $\mathcal{D}$, the BCS result is only valid when the size of $\mathcal{D}$ is small relative to the Fermi level.
The DFM was introduced to overcome this limitation in the weak-coupling regime \cite{kirzhnits73}.

The normalized gap function $\phi(\omega) = \Delta(\omega)/\Delta(0)$ can be determined from Eqs.~(\ref{eq:gapequation})-(\ref{eq:kernel}) using Zubarev's~\cite{zubarev60} approach.
This method is valid for low temperatures, where the Fermi level is much larger than the thermal energy of the charge carriers.
The normalized gap function is then given by the (numerical) solution to the following Fredholm equation of the second kind:
\begin{equation}
	\label{eq:phizub}
	\phi(\omega) = \frac{K(\omega,0)}{K(0,0)} - \int \limits_{-\epsilon_F}^\infty \frac{\dd\omega'}{2|\omega'|} \phi(\omega')\left[ K(\omega,\omega') - \frac{K(0,\omega') K(\omega,0)}{K(0,0)} \right].
\end{equation}
Once the normalized gap function has been determined, the critical temperature is found by
\begin{equation}
	\label{eq:tczub}
	T_c = \frac{2 e^\gamma}{\pi} \epsilon_F \exp \left( -\frac{1}{\lambda} \right),
\end{equation}
where $\gamma \approx 0.5772\dots$ is the Euler-Mascheroni constant and the parameter $\lambda$ is defined as
\begin{equation}
	-\frac{1}{\lambda} =  \frac{1}{K(0,0)} + \int_{-\epsilon_F}^\infty \frac{\dd \omega}{2|\omega|} \left[ \frac{K(0,\omega)\phi(\omega)}{K(0,0)} - \Theta(\epsilon_F-|\omega|) \right],
\end{equation}
with $\Theta(x)$ the Heaviside step function. 

\subsection{\label{subsec:dielec}Dielectric function}
In this subsection, the interaction potential of our model is determined.
In the weak-coupling regime, the inverse of the dielectric function describes the response of the system to an external perturbation.
Thus, the function
\begin{equation}
	V(\bq, i\Omega) = \frac{2\pi e^2}{|\bq| \epsilon(\bq, i\Omega)}
\end{equation}
characterizes the interelectron effective potential as a screened Coulomb potential, with the electron charge $e$ and the dielectric function $\epsilon(\bq, i\Omega)$ which contains all screening effects.
Within RPA, the dielectric function is given by Lindhard's formula\cite{pines94}
\begin{equation}
	\epsilon(\bq, i\Omega) = 1 - \frac{2\pi e^2}{|\bq|}\chi (\bq, i\Omega).
	\label{eq:dielectric}
\end{equation}
The graphene density-density response function $\chi(\bq,i\Omega)$ is \cite{wunsch06}
\begin{equation}
	\chi(\bq, i\Omega) = 4\sum_{\lambda,\lambda',\bk} |\mathcal{D}_{\lambda\lambda'}(\bk, \bk+\bq)|^2
					\frac{n_{\lambda,\bk} - n_{\lambda',\bk + \bq}}{i\Omega + \epsilon_{\lambda,\bk} - \epsilon_{\lambda',\bk+\bq}},
\end{equation}
with $n_{\lambda,\bk}$ the Fermi-Dirac distribution and the density vertex $|\mathcal{D}_{\lambda\lambda'}(\bk, \bk+\bq)|^2 $ determined as
\begin{equation*}
	\begin{aligned}
		|\mathcal{D}_{\lambda\lambda'}(\bk, \bk+\bq)|^2 &= |\braket{\Psi_{\lambda,\bk }|\Psi_{\lambda',\bk+\bq } }|^2\\
		&= \frac{1}{2}\left[(1 + \lambda\lambda'\cos(\Delta\phi)\right].
	\end{aligned}
	\label{eq:densvertex}
\end{equation*}
The angle $\Delta\phi$ is the angle between the vectors $\bk$ and $\bk + \bq$.
In the low-temperature regime where Zubarev's approach is valid, the Fermi-Dirac distribution is well represented by a Heaviside step function. 
This way, the summations over $\lambda$ and $\lambda'$ can be carried out. 
The summation over $\bk$ can be converted to an integration, which gives
\begin{widetext}
\begin{equation}
	\begin{aligned}
		\chi(\bq,i\Omega) = 
		&-\frac{2 k_F}{\pi^2 v_F}\int \limits_0^1\dd k \int \limits_{|k-q|}^{k+q} \dd y \frac{2ky + k^2 + y^2 - q^2}{\sqrt{4k^2q^2 - (y^2-k^2-q^2)^2}} 
		\frac{y - k }{\Omega^2 + ( y - k )^2}\\
		&-\frac{2 k_F}{\pi^2 v_F} \int \limits_1^\frac{\Lambda}{k_F} \dd k \int \limits_{|k-q|}^{k+q}\dd y \frac{2ky - k^2 - y^2 + q^2}{\sqrt{4k^2q^2 - (y^2-k^2-q^2)^2}} 
		\frac{y + k}{\Omega^2+ ( y + k )^2 }.
	\end{aligned}
	\label{eq:pol}
\end{equation}
\end{widetext}
The variables $\bq$ and $\Omega$ have been made dimensionless by the substitutions $\bq \rightarrow k_F \bq$ and $\Omega \rightarrow \epsilon_F\Omega$. 
The integration variables $k$ and $y$ are also made dimensionless. 
The parameter $\Lambda$ is a cutoff wavevector, stemming from the discreteness of the graphene lattice and is set to $\Lambda \approx 8$ eV \cite{peres06}.

Screening effects by the lattice are included as harmonic phonon contributions in the dielectric function obtained within RPA
\begin{equation}
	\epsilon(\bq, i\Omega) = \epsilon_L(\bq, i\Omega) - \frac{2\pi e^2}{|\bq|}\chi (\bq, i\Omega),
	\label{eq:dielectricfull}
\end{equation}
where the lattice dielectric function is
\begin{equation*}
	\epsilon_L(\bq, i\Omega) = \kappa \left(\frac{\Omega^2 + \omega_{LO}^2(q)}{\Omega^2 + \omega_{TO}^2(q)}\right).
\end{equation*}
The longitudinal and transverse optical phonon dispersions, $\omega_{LO}(q)$ and $\omega_{TO}(q)$ respectively, were previously calculated for graphene by Maultzsch \textit{et al.}~\cite{maultzsch04} and Mounet and Marzari~\cite{mounet05}. 
The high-frequency dielectric constant $\kappa$ depends on the environment of the graphene sheet. 
An isolated sheet of graphene has the dielectric constant $\kappa = 1$. 
However, by placing graphene on top of, for example, a layer of hexagonal Boron nitride (hBN) with $\kappa_{\textrm{hBN}} = 3$ \cite{young12}, the dielectric constant can be increased.

Phonon mediated pairing of electrons at low carrier doping will rely on phonons in the low momentum regime around the $\Gamma$-point of the phonon dispersion.
Since the LO and TO branches coincide for $|\bq| \approx 0$, the lattice dielectric function will be $\epsilon_L \approx \kappa$.
Visible in Mounet and Marzari's calculation of the phonon dispersion, there is a third optical branch, consisting of flexural out-of-plane phonons.
As discussed by Mariani and von Oppen~\cite{mariani08}, these flexural phonons contribute only at higher orders of perturbation, since they  couple to charge carriers through a two-phonon vertex, due to their reflection symmetry.
Therefore, we choose to neglect the contribution of flexural phonons in the electron-phonon pairing mechanism.
Thus, the dielectric function used in our calculations will be
\begin{equation}
	\epsilon(\bq, i\Omega) = \kappa - \frac{2\pi e^2}{|\bq|}\chi (\bq, i\Omega),
\end{equation}
with the dielectric constant $\kappa$ as a parameter indicative of the immediate environment of the graphene sheet.

\section{\label{sec:results}Results and discussion}
To obtain the gap function and critical temperature given by Eqs.~(\ref{eq:phizub}) and (\ref{eq:tczub}), only several material parameters are needed: the C-C bond length $a \approx 1.42$ \AA , the hopping parameter $t \approx 2.8$ eV, and the discrete lattice parameter $\Lambda \approx 8$ eV. 
The final parameter needed is the charge carrier density, which we vary between $10^{10}$~cm$^{-2}$ and $10^{12}$~cm$^{-2}$. 
The upper limit of this range is determined by the validity of the Dirac cone for the dispersion relation. 
For energies above $\approx 0.1$ eV, the Dirac cone is no longer a valid approximation of the graphene band structure.
This energy corresponds with the charge carrier density $n = 10^{12}$~cm$^{-2}$. 
Local variations in electron or hole doping, so called electron-hole puddles, create the lower boundary for the charge carrier densities treated in this work \cite{martin08}.

\begin{figure}[htb]
	\centering
    \includegraphics[keepaspectratio=true,width=80mm]{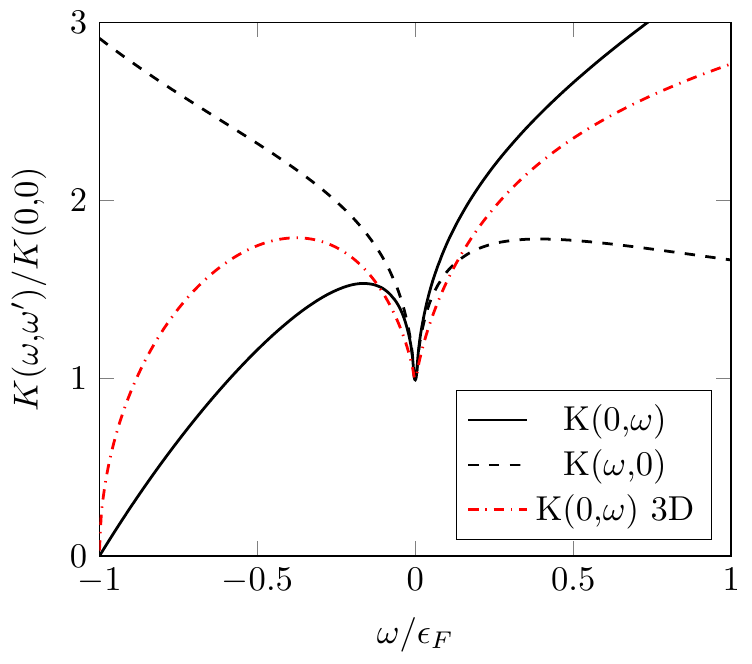}
	\caption{(Color online) Solid and dashed curves: the $K(0,\omega)$ (solid) and $K(\omega,0)$ (dashed) functions, calculated at a charge carrier density of $10^{10}$~cm$^{-2}$. Dash-dotted curve: the typical shape of the $K(0,\omega)$ function for a 3D system with parabolic dispersion.}
	\label{fig:kernel}
\end{figure}

The solid and dashed curves in Fig.~\ref{fig:kernel} show the kernel for graphene along the two axes in $(\omega,\omega')$-space.
Both curves exhibit a local minimum at the origin $\omega,\omega' = 0$.
There, the shape of the kernel is determined by both the attractive electron-plasmon interaction and the Coulomb repulsion.
Away from the Fermi surface, the kernel is dominated by the Coulomb repulsion.
This way, Cooper pair formation takes place predominantly in the region of attraction around the Fermi surface.
For $\omega \to -\epsilon_F$, the graphene kernel tends to zero due to the prefactor $|\bk|$ in Eq.~\ref{eq:kernel}.
This prefactor is a consequence of the linear dispersion of the Dirac cone.
The kernel of a 2D system with a parabolic energy dispersion attains a finite value for $\omega = -\epsilon_F$ and monotonically decreases towards the local minimum in the origin~\cite{takada78,klimin14}.
In this regard, the graphene kernel resembles more the kernel of a 3D system with parabolic dispersion, illustrated by the dash-dotted line in Fig.~\ref{fig:kernel}.

\begin{figure}[hbt]
	\centering
	\includegraphics[keepaspectratio=true,width=80mm]{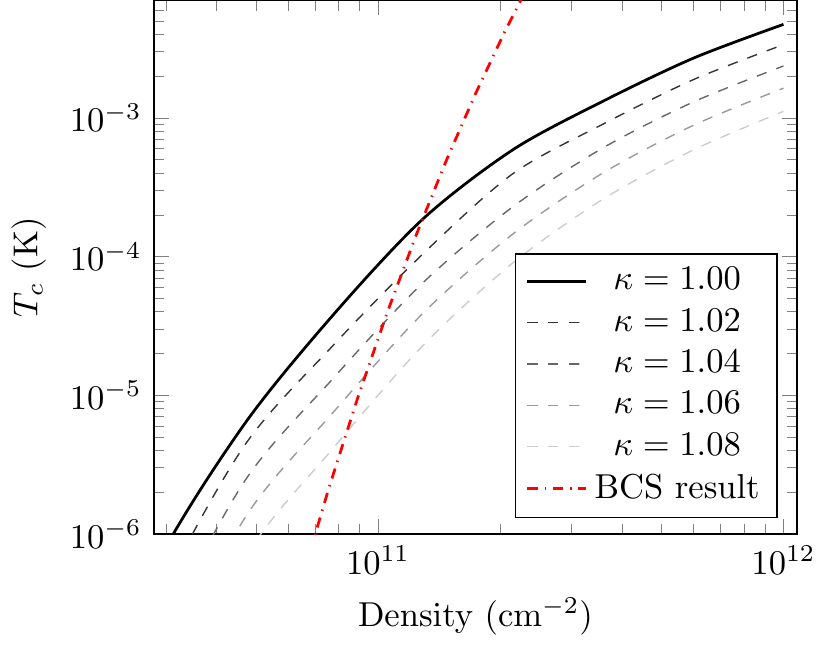}
	\caption{(Color online) Solid and dashed curves: critical temperatures as a function of carrier density, for a range of dielectric constants $\kappa$. Lines become lighter for increasing values of $\kappa$. Dash-dotted curve: the BCS result by Kopnin and Sonin~\cite{kopnin08} for a coupling constant of 0.01 and a Debye window of 5 meV.}
	\label{fig:kappa}
\end{figure}

The solid curve in Fig.~\ref{fig:kappa} shows the critical temperature for the range of electron densities of $10^{10}$~cm$^{-2}$ -- $10^{12}$~cm$^{-2}$ with a dielectric constant of $\kappa = 1$. 
For low densities, the critical temperature is strongly suppressed due to the low density of states around the Dirac point. 
The critical temperature rises monotonically to the millikelvin range for increasing electron densities.
When additional screening is present ($\kappa > 1$), the critical temperature diminishes over the entire electron density range, as indicated by the dashed curves.
While the plasmon mediated attraction relies on a second order (screened) Coulomb interaction (electron - plasmon - electron), the Coulomb repulsion is first order.
Thus, for increased screening, Cooper pair formation becomes more difficult, decreasing the critical temperature.

The critical temperatures obtained here are small (in the millikelvin or microkelvin regime) and display a more gentle dependence on the doping density when compared to the BCS result~\cite{kopnin08} illustrated by the dash-dotted curve in Fig.~\ref{fig:kappa}.
However, the BCS critical temperatures are highest for an interaction energy range that is comparable to the Fermi energy.
This violates one of the basic BCS assumptions, namely a small Debye window with respect to the Fermi level. 
Hence, we expect the current approach to be more appropriate than the standard BCS approach. 

Finally, note that for two-dimensional systems, superconducting coherence is lost via the Berezinskii-Kosterlitz-Thouless (BKT) mechanism~\cite{berezinskii71,kosterlitz72}, rather than by pair breaking excitations.
The critical temperature computed here relates to the pair-breaking gap and does not capture the BKT mechanism.
However, even though the critical temperatures predicted here only indicate an upper limit for the BKT transition temperature, this should not be a major issue in the weak-coupling regime.
The BKT transition temperature $T_{\textrm{BKT}}$ is proportional to the superfluid density.
In the weak-coupling regime, the thermal energy at the BKT transition temperature is small with respect to the Fermi energy.
Thus at this temperature, the superfluid density is small with respect to the total density.
Since the superfluid density decreases when $T$ approaches $T_c$, at weak coupling $T_{\textrm{BKT}}\approx T_c$.

\section{\label{sec:conclusions}Conclusions}
We derived the relevant equations using DFM for the critical temperature of superconductivity in graphene. 
Compared to previous results, the kernel shape is more similar to DFM results of 3D systems with parabolic dispersion than to the results of other 2D systems, due to graphene's linear band dispersion. 
The calculated critical temperatures are suppressed for low doping densities, as expected. 
The transition temperature of plasmon-mediated Cooper pairing in graphene also decreases for increasing dielectric constants, due to increased screening of the Coulomb potential.
The discrepancy with the BCS results shows that also for future applications that use bilayers or rely on the dielectric environment of graphene to boost and probe superconductivity, the dielectric function method is more appropriate than the standard BCS or Eliashberg approach.

\section{\label{sec:acknowledgements}Acknowledgements}
This research was supported by the joint FWO-FWF project POLOX (Grant No. I 2460-N36) and the Bijzonder Onderzoeksfonds (BOF) of the Research Council of the University of Antwerp.

\bibliography{bibDFMGraphene}

\end{document}